# Metabolically Stable Neurotensin Analogs Exert Potent and Long-Acting Analgesia Without Hypothermia

Mélanie Vivancos[a,b], Roberto Fanelli[c], Élie Besserer-Offroy[a,b,1], Sabrina Beaulieu[a,b], Magali Chartier[a,b], Martin Resua-Rojas[a,b], Christine E. Mona[d], Santo Previti[c,2], Emmanuelle Rémond[c], Jean-Michel Longpré[a,b], Florine Cavelier[c,3,*], Philippe Sarret[a,b,3,**]

[a]Department of Pharmacology-Physiology, Faculty of Medicine and Health Sciences, Université de Sherbrooke, Sherbrooke, Québec, Canada

[b]Institut de Pharmacologie de Sherbrooke, Université de Sherbrooke, Sherbrooke, Québec, Canada

[c]Institut des Biomolécules Max Mousseron (IBMM), UMR-CNRS 5247, Université Montpellier, ENSCM, Montpellier, France

[d]Department of Molecular and Medical Pharmacology, David Geffen School of Medicine at the University of California at Los Angeles, Los Angeles, CA, USA

[1]Present address: Department of Molecular and Medical Pharmacology, David Geffen School of Medicine at the University of California at Los Angeles, Los Angeles, CA, USA

[2]Present address: Department of Chemical, Biological, Pharmaceutical and Environmental Sciences, University of Messina, Vial Annunziata, Messina, Italy

[3]Lead Author

**Corresponding Authors:**

[**]**Philippe Sarret, Ph.D.**
Dept of Pharmacology-Physiology
Université de Sherbrooke
3001, 12e Avenue Nord
Sherbrooke, QC, Canada, J1H 5N4
Philippe.Sarret@USherbrooke.ca
Phone: +1 (819) 821-8000 ext. 72554

[*]**Florine Cavelier, Ph.D.**
IBMM, UMR-CNRS-5247
Université de Montpellier
19 Place Eugène Bataillon
34095 Montpellier Cedex 5, France
Florine.Cavelier@umontpellier.fr
Phone: +33 (0) 467 14 37 65

**Email addresses of authors:**

melanie.vivancos@usherbrooke.ca (MV)

robyfanelli@hotmail.it (RF)

sabrina.beaulieu@usherbrooke.ca (SB)

martin.resua-rojas@usherbrooke.ca (MRR)

cmona@mednet.ucla.edu (CEM)

elie.besserer@mcgill.ca (ÉBO)

emmanuelle.remond@umontpellier.fr (ER)

spreviti@unime.it (SP)

jean-michel.longpre@usherbrooke.ca (JML)

florine.cavelier@umontpellier.fr (FC)

philippe.sarret@usherbrooke.ca (PS)

**Highlights**

- Incorporation of reduced amide bond, silaproline and TMSAla in NT(8-13) resulted in JMV5296, a 25-fold NTS2-selective analog.

- Combination of these three chemical modifications increased resistance toward proteases having a plasma stability over 20 hours.

- Intrathecal injection of JMV5296 induced potent antinociception in acute, tonic and chronic inflammatory pain models.

- Central delivery of JMV5296 had no impact on body temperature.

**Chemical compounds studied in this article**

NT(8-13) (PubChem CID 5311318)

PD149163 (PubChem CID 73064239)

JMV431 (Pubchem SID 135652223)

**Abstract**

The endogenous tridecapeptide neurotensin (NT) has emerged as an important inhibitory modulator of pain transmission, exerting its analgesic action through the activation of the G protein-coupled receptors, NTS1 and NTS2. Whereas both NT receptors mediate the analgesic effects of NT, NTS1 activation also produces hypotension and hypothermia, which may represent obstacles for the development of new pain medications. In the present study, we implemented various chemical strategies to improve the metabolic stability of the biologically active fragment NT(8-13) and assessed their NTS1/NTS2 relative binding affinities. We then determined their ability to reduce the nociceptive behaviors in acute, tonic, and chronic pain models and to modulate blood pressure and body temperature. To this end, we synthesized a series of NT(8-13) analogs carrying a reduced amide bond at $Lys^8$-$Lys^9$ and harboring site-selective modifications with unnatural amino acids, such as silaproline (Sip) and trimethylsilylalanine (TMSAla). Incorporation of Sip and TMSAla respectively in positions 10 and 13 of NT(8-13) combined with the $Lys^8$-$Lys^9$ reduced amine bond (JMV5296) greatly prolonged the plasma half-life time over 20 hours. These modifications also led to a 25-fold peptide selectivity toward NTS2. More importantly, central delivery of JMV5296 was able to induce a strong antinociceptive effect in acute (tail-flick), tonic (formalin), and chronic inflammatory (CFA) pain models without inducing hypothermia. Altogether, these results demonstrate that the chemically-modified NT(8-13) analog JMV5296 exhibits a better therapeutic profile and may thus represent a promising avenue to guide the development of new stable NT agonists and improve pain management.

1. Introduction

Chronic pain is an important public health problem affecting more than 20% of the worldwide population [1]. Opioid drugs are extensively used in the treatment of chronic pain despite a long list of undesired effects and their limited long-term efficacy to relieve pain for many patients [2, 3]. Even with the growing awareness of the risks associated with opioid misuse, overdose and addiction, opioid use is still rising, thus leading to the current opioid crisis in North America [4, 5]. Thus, the societal and economic burden, healthcare costs and high prevalence of chronic pain encourage researchers to seek for new pain medications with an increased benefit/risk ratio [6, 7]. Among the current development strategies, drugs targeting non-opioid G protein-coupled receptors (GPCR) represent a promising therapeutic avenue in pain research [7].

The development of peptide-based therapeutics is undergoing an exciting revival in the last decade, when compared to small molecule drugs [8, 9]. Peptides often offer high target selectivity and specificity as well as enhanced efficacy, safety and tolerability profiles. However, naturally occurring peptides are often not directly suitable for clinical use due to low oral bioavailability, poor blood-brain barrier (BBB) penetration, and short half-life in physiological fluids related to their poor resistance to proteolytic degradation [8, 10].

Among the promising alternatives to opioid pain medications, neurotensin (NT) receptors emerge as attractive targets for the treatment of pain [7, 11]. Neurotensin (NT) is a small neuropeptide of 13 amino acids (pGlu-Leu-Tyr-Glu-Asn-Lys-Pro-Arg-Arg-Pro-Tyr-Ile-Leu) [12] known to mediate its physiological effects through its binding to two receptors that belongs to 7TMRs superfamily, namely NTS1 and NTS2 [13]. Peripherally, NT acts as a hormone in the cardiovascular system where it induces a drop in blood pressure [14, 15], controls appetite [16] and regulates gastrointestinal motility [17]. When administered directly into the central nervous system (CNS), NT is known to play a role in the regulation of anxiety [18, 19] as well as in dopaminergic (DAergic)-associated diseases, such as schizophrenia, drug abuse, and Parkinson's disease [20-22]. NT and its analogs also produce persistent hypothermia [20, 23] and analgesia [24-26]. Both NTS1 and NTS2 receptors are present in CNS regions involved in nociceptive transmission and pain modulation, such as the spinal dorsal horn, periaqueductal gray (PAG), rostroventral medulla (RVM), dorsal raphe nucleus, raphe magnus and pallidus [25, 27-29]. Moreover, the neurotensinergic system is gaining further interests for pain relief as NT-induced analgesia is not altered by the administration of the opioid antagonists' naloxone and naltrexone, thereby supporting that NT receptor activation mediates its antinociception action independently of the opioid system [30].

Amino acids forming the C-terminal moiety of the native NT peptide, at position 8 to 13 (Arg-Arg-Pro-Tyr-Ile-Leu), were identified as the minimal sequence for NT receptor binding and biological activity [31, 32]. Therefore, compound synthesis and structure-activity relationship studies have allowed the development of new NT(8-13) selective analogs targeting either NTS1 or NTS2 receptors and permitted the discrimination of their respective physiological effects [33, 34]. While analgesia in acute, tonic, and chronic pain paradigms was demonstrated to be mediated by the activation of both receptors [24, 25, 28, 34-36], only NTS1 activation was associated with hypotension [14, 37] and hypothermia [38].

These adverse effects (hypothermia and hypotension) combined with the low metabolic stability of the native NT peptide preclude its therapeutic use as non-opioid pain-relieving medications. Therefore, chemical modifications of the backbone and the incorporation of unnatural amino acids are necessary to optimize the pharmacological properties of newly synthesized NT(8-13) drug candidates. In the present study, we evaluate the impact of site-selective chemical modifications of NT(8-13) to improve the peptide metabolic stability and its analgesic efficacy in different experimental pain models as well as to reduce the unwanted effects triggered by NTS1 activation.

## 2. Materials and Methods

*2.1 Peptide chemistry and characterization*

Full synthetic procedures and characterization of the compounds presented in this study are reported in [39].

*2.2 Competitive radioligand binding assay*

CHO-K1 cells stably expressing hNTS1 (ES-690-C from PerkinElmer, Montréal, Canada) or 1321N1 cells stably expressing hNTS2 (ES-691-C from PerkinElmer) were cultured respectively in DMEM/F12 or DMEM. Culture media were supplemented with 10% FBS, 100 U/mL penicillin, 100 μg/mL streptomycin, 20 mM HEPES and 0.4 mg/mL G418, and cells were incubated at 37°C in a humidified chamber at 5% $CO_2$. All media and additives are from Wisent (St-Bruno, Canada). Competitive radioligand binding experiments were performed as previously described [40]. Briefly, 50 μg of cell membranes, expressing either hNTS1 or hNTS2, were incubated with $^{125}I$-$Tyr^3$-NT (2200 Ci/mmol, from PerkinElmer, Billerica, MA). Competition was done with increasing concentrations of NT analogs diluted in binding buffer (50 mM Tris-HCl, pH 7.5, 0.2 % BSA) and ranging from $10^{-11}$ to $10^{-5}$ M. After one hour of incubation at room temperature, the binding reaction was terminated by filtration in 96-well glass-fiber filter plates (GF/C, Millipore, Billerica, MA) and plate was washed three times with ice-cold binding buffer. Filters were then counted in a γ-counter (1470 Wizard2, PerkinElmer). Non-specific binding was measured in the presence of $10^{-5}$ M unlabeled NT(8–13) and represented less than 5% of total binding. $IC_{50}$ values were determined from the competition curves as the unlabeled ligand concentration inhibiting half of the $^{125}I$-$Tyr^3$-NT-specific binding. Data were plotted using

GraphPad Prism 8 using the One-site – Fit log (IC$_{50}$) and represent the mean ± SEM of at least three separate experiments.

IC$_{50}$ calculated from the competitive radioligand binding assays were then transformed into K$_i$ using the Cheng−Prusoff equation [41] :

$$K_i = \frac{IC_{50}}{\left(1 + \frac{[L]}{K_d}\right)}$$

where L refers to the concentration of radiolabeled tracer ($^{125}$I-[Tyr$^3$]- NT) and K$_d$ refers to the equilibrium dissociation constant of the radioligand. Determined K$_d$ are 0.7 nM and 3.4 nM for NTS1 and NTS2, respectively.

## 2.3 Plasma stability

Rat plasma was obtained from blood by keeping the translucent phase after centrifugation at 15,000g over 5 min. Plasma stability assay was carried out by incubating compounds in rat plasma at a final concentration of 0.156 mM. NT(8-13), JMV438, and JMV449 were incubated 0, 1, 2, 5, 10, 30 and 60 min, whereas JMV5206, JMV5170, and JMV5296 were incubated for longer times (0, 1, 2, 4, 8, 16 and 24 hours) at 37°C. Then, 70 µl of a solution containing 10% Trichloroacetic acid (TCA) and 0.5% nicotinamine, as an internal standard, was added to stop the degradation by proteases. After centrifugation at 15,000g for 30 min, supernatant was filtered through 0.22 µm filter and analyzed by UPLC/MS for quantification, as previously described [42]. % of remaining peptide over time was graphed into GraphPad Prism 8 and the half-life of each compound was calculated using one-phase decay fit and represented the mean ± SEM of at least three separate experiments.

## 2.4 Animals, housing and habituation

Experimental procedures were approved by the Animal Care Committee of the Université de Sherbrooke (protocol n°035-18B) and were in accordance with policies and directives of the *Canadian Council on Animal Care*. Furthermore, all procedures involving animals followed the ARRIVE recommendations [43].

Adult male Sprague-Dawley rats, weighing 200-225 g (Charles River laboratories, St-Constant, Québec, Canada) were maintained on a 12 h light/dark cycle with free access to food and water. Rats were housed two by transparent open-top cage on aspen shavings in a quiet room. Seven days upon arrival, animals were acclimatized for 4 days to the experimental room and for 3 days to the manipulations and testing devices prior to the behavioral studies. Sex was not taken into account in the present study.

### *2.4.1 Behavioral studies*

Behavioral experiments were performed by three female experimenters in a quiet room between 08.00 AM and 12.00 PM to reduce variation related to the circadian rhythm.

### *2.4.2 Intrathecal administration of compounds*

Rats randomly assigned to control and experimental groups were lightly anesthetized with 2.5% isoflurane/oxygen (Baxter corporation, Mississauga, ON, Canada; 2 L/min) and injected intrathecally (i.t.) with a 27 G1/2 needle in the L5−L6 intervertebral space with 0.9% saline or 25 µl of each compound, as previously published [44-46]. All NT(8-13) analogs were diluted in saline at 5 mg/ml.

*2.4.3 Tail-flick test*

Effects of i.t. injection of NT analogs, vehicle or morphine at equimolar dose were assessed on acute thermal nociception using the tail-flick test (Tail-Flick Analgesia meter V2.00, Columbus Instruments, Columbus, Ohio, USA). Tail-flick test measures sensitivity to a high-intensity light beam focused on the tail of the rat. The tail withdrawal latency (time to curl or flick tail out of light beam path, in s), relates to pain sensitivity. Light intensity was set at 6 and a determined 10 s cut-off was used. On the test day, baseline latencies measures were taken before drug injection to provide a mean baseline. Tail-flick latencies were measured at baseline and at each 10 min for up to 40 min after drug administration. Data are expressed as mean ± SEM for 5 – 8 rats in each compound-treated group.

*2.4.4 Formalin test*

The analgesic effects of NT(8-13), JMV5296 and morphine were assessed in a tonic pain paradigm using the formalin test. Five minutes after i.t. injection of saline, NT(8-13) or JMV5296 (60 µg/kg), or equimolar dose of morphine, rats received a 50 µl of diluted 2% formaldehyde (i.e. 5% formalin; Bioshop, Burlington, Canada) into the plantar surface of the right hind paw. Immediately, rats were placed in clear plastic chambers (30 × 30 × 30 cm) positioned over a mirror angled at 45° to allow an unobstructed view of the paws and their behaviors were observed for the next 60 min. An intra-plantar injection of formalin produced the biphasic nocifensive response typical of this persistent pain model [47]. The two distinct phases of spontaneous pain behaviors that occur in rodents are proposed to reflect a direct effect of formalin on sensory receptors (acute phase, 0-9 min post-intraplantar injection) and a longer lasting pain due to inflammation and central sensitization (inflammatory phase, 21-60 min post-intraplantar injection).

Nociceptive behaviors were assessed using a weighted score as described previously [48, 49]. Following injection of formalin into the hind paw, nociceptive mean score was determined for each 3 min block during 60 min by measuring the time spent in each of four behavioral categories: 0, the injected paw is comparable to the contralateral paw; 1, the injected paw has little, or no weight placed on it; 2, the injected paw is elevated and is not in contact with any surface; 3, the injected paw is licked, bitten, or shaken [48]. The behaviors believed to represent higher levels of pain intensity were given higher weighted scores. The weighted average pain intensity score ranging from 0 to 3 was then calculated by multiplying the time spent in each category by the category weight, summing these products, and dividing by the total time in a given time interval. The pain score was thus calculated from the following formula $(1 \times T1 + 2 \times T2 + 3 \times T3)/180$ where $T1$, $T2$, and $T3$ are the durations (in seconds) spent in behavioral categories 1, 2, or 3, respectively, during each 180 second block. The area under the curve (AUC) was calculated during all the duration of the test (0-60 min) using GraphPad Prism 8. Data represent the mean ± SEM of 5 animals for each condition.

### *2.4.5 Chronic inflammatory pain (CFA)*

Chronic inflammatory pain was induced with an intraplantar injection of 100 μl of Complete Freund's Adjuvant (CFA) (Calbiochem, USA) using a 25 G needle. CFA was injected in the plantar surface of the left hind paw of rats under isoflurane anesthesia. The injected CFA solution was supplemented with desiccated Mycobacterium butyricum (Becton, Dickinson and company, Sparks, USA) for a final concentration of 4 mg of bacteria membrane/ml. Saline was added to the mixture to prepare a homogenous 1:1 (w:w) emulsion.

To determine the presence of mechanical allodynia, rats were placed in enclosures with an elevated wire mesh floor. A series of 5 von Frey hairs (2, 4, 6, 10, 15 g) was applied alternately on both ipsilateral and contralateral hind paws at 15 s intervals to measure tactile allodynia. The von Frey filament was directed to the plantar surface with sufficient force to cause slight curvature of the hair [50]. A positive response was recorded when the paw was withdrawn, or the pre-set cut-off was reached (15 g). Measurements of mechanical allodynia were performed before injection of CFA (baseline, day 0). Then, 3 and 14 days after induction of inflammatory pain, NT(8-13), JMV5296 or saline were injected i.t. and the presence of mechanical allodynia was assessed. Furthermore, percentage of anti-allodynic effect was determined at the analgesic peak and calculated using the following equation: % Anti-allodynic effect = 100 × [(CFA) – (baseline)] / [(sham) – (baseline)]. From the latter formula, 0% represents no anti-allodynic effect of the compound, while 100% corresponds to a complete relief of mechanical hypersensitivity. Data represent the mean ± SEM of 7 – 8 animals for each condition.

### 2.4.6 Core body temperature

Body temperature was measured using a thermistor probe inserted into the rectum of adult Sprague-Dawley rats. Temperatures were recorded immediately before administration of compounds (baseline) and each 10 min for up to 60 min following i.t. administration of saline or NT analogs at 30 μg/kg. Changes in body temperature from baseline (Δ body temp) were determined for each animal. Data are expressed as mean ± SEM, of at least 5 animals.

### 2.4.7 Blood pressure monitoring

Rats were anesthetized with a mixture of ketamine/xylazine (87 mg/kg: 13 mg/ kg, i.m,) and placed in a supine position on a heating pad. Mean, systolic and diastolic arterial blood pressure as well as heart rate were measured through a catheter (PE 50 filled with heparinized saline) inserted in the right carotid artery and connected to a Micro-Med transducer (model TDX-300,

USA) linked to a blood pressure Micro-Med analyzer (model BPA-100c). Another catheter (PE 10 filled with heparinized saline) was inserted in the left jugular vein for injections (1 mL/kg, 5–10 sec) of saline or NT analogs at 0.01 mg/kg. Blood pressure was recorded each 5 sec for up to 300 sec following intravenous injection. For relative potency evaluation, changes in arterial blood pressure ($\Delta$ MABP) were determined from the basal pressure of the rat. Data represents the mean ± SEM of 5 – 10 animals for each condition.

*2.5 Statistical analysis*

All graphs and statistical analysis were performed using GraphPad Prism 8 (GraphPad software, La Jolla, CA, USA). Normality of variables was assessed for all *in vivo* related results prior to the application of any other statistical tests using the Shapiro-Wilk test (alpha, 0.05). A two-way ANOVA followed by Bonferroni's test in multiple comparisons was used to determine significant differences between drug- and saline-treated rats in tail-flick, body temperature, formalin, and CFA-treated rats. For AUC in tail-flick, formalin test and percentage of % of anti-allodynic effect, a one-way ANOVA followed by Dunnett's post hoc multiple comparison was used to compare drugs and saline. A difference in response between saline and NT analogs was considered significant with a *p*-value < 0.05.

3. **Results and discussion**

*3.1 Rational design of NT(8-13) analogs*

Both Arg residues in positions 8 and 9 were replaced by two Lys for ease synthesis (JMV438). It was previously demonstrated that this amino acid di-substitution had no impact on NTS1 and NTS2 receptor binding or activation [51-53]. Therefore, the subsequent chemical modifications of the NT(8-13) sequence were based on the JMV438 derivative.

First, the chemical modifications performed on the NT(8-13) backbone were focused on increasing the metabolic stability of NT(8-13) analogs to proteolytic cleavage at $Arg^8$-$Arg^9$ and $Pro^{10}$-$Tyr^{11}$ by three metalloendopeptidases referred to as EC 3.4.24.11 (also known as nephrilysin), EC 3.4.24.15 (thimet oligopeptidase), and EC 3.4.24.16 (neurolysin)[54-56]. Therefore, $Arg^8$-$Arg^9$ was replaced by a reduced amine bond ($\Psi[CH_2NH]$) between $Lys^8$-$Lys^9$ (JMV449), this N-terminal protection conferring resistance against proteolytic degradation, with minor impact on NT receptor binding [39, 51, 52]. On JMV5170, $Pro^{10}$ was then substituted by the silylated proline analog silaproline (Sip) [57]. This isostere unnatural amino acid of proline introduced at position 10 protects against the second enzymatic cleavage. We recently demonstrated that the replacement of $Pro^{10}$ by $Sip^{10}$ did not affect the compound's structure [57] or receptor binding affinity, however, this modification slightly increased its resistance toward proteases [36, 58].

Secondly, the C-terminal end of NT(8-13) is constituted of lipophilic residues $Ile^{12}$ and $Leu^{13}$ that play an important role for the recognition and activation of NT receptors [59, 60]. It was previously reported that the substitution of $Leu^{13}$ by the (L)-(Trimethylsilyl)-Alanine (TMSAla) residue increased affinity toward both NTS1 and NTS2 receptors [61]. From the aforementioned modifications, a series of NT(8-13) analogs harboring a reduced amine bond combined with a TMSAla (JMV5206) or a di-substitution with both Sip and TMSAla (JMV5296) were synthetized and characterized *in vitro* for their binding affinities and plasma stability (**Table 1**). We then assessed the influence of these chemical modifications on the analgesic properties of these NT(8-13) analogs in preclinical models of acute, tonic and inflammatory chronic pain and their ability to induce NTS1-mediated hypotension and hypothermia.

*3.2 Site-selective modifications with unnatural amino acids generate stable and selective NT(8-13) analogs.*

Binding affinities of NT(8-13) analogs were carried out on membranes prepared from CHO-K1 cells stably expressing hNTS1 or 1321N1 cells stably expressing hNTS2. The *in vitro* stability of these compounds was determined in rat plasma and their degradation profiles assessed by UPLC-MS. Receptor binding affinity, selectivity, as well as peptide half-life in plasma are summarized in **Table 1**. We observed that the substitution of the N-terminal Arg doublet of NT(8-13) by Lys-Lys (JMV438) did not impact NTS1 and NTS2 binding, as it was previously reported [51-53]. The incorporation of a reduced amine bond ($\Psi[CH_2NH]$) between $Lys^8$-$Lys^9$ (JMV449) induced a slight increase in binding affinity at both NTS1 and NTS2 receptors, thus creating a compound with a small selectivity gain toward NTS2 of nearly 10-fold when compared to the native NT(8-13). Moreover, the presence of the reduced bond slightly improved the plasma stability by a factor of 8.5 in comparison with NT(8-13). This chemical modification was then retained in all the synthesized compounds. Substitution of $Pro^{10}$ by Sip (JMV5170) induced a loss of binding affinity to NTS1 and NTS2 receptors, when compared to JMV449 (~150- and ~450-fold decrease for NTS1 and NTS2, respectively). However, this substitution greatly improved the degradation profile of JMV5170 with a more than 150-fold increase in plasma half-life (> 20 hours), compared to JMV449 (< 10 min). This was previously observed, in a lesser extent, with a NT(8-13) analog bearing a Sip in position 10 but lacking the reduced amine bond between $Lys^8$-$Lys^9$ (namely, JMV2009; H-Lys-Lys-Sip-Tyr-Ile-Leu-OH) [36, 58]. Indeed, JMV2009 displayed a slightly reduced affinity at both NTS1 and NTS2 receptor sites but an extended *ex-vivo* plasma half-life when compared to the full-length NT peptide, undoubtedly mediated by a reduced recognition by metallo-endopeptidases. The incorporation of a TMSAla residue in position 13, together with the reduced amine bond at the N-terminal dibasic residues (JMV5206) did not affect the binding affinities for NTS1 and NTS2 compared to JMV449 but led to a moderate increase in plasma stability with a half-life of more than 2 hours. Interestingly, this combination did not improve binding affinities toward NT receptors as we would have expected when looking at its counterpart without the reduced bond, namely JMV2007 (H-Lys-Lys-Pro-Tyr-Ile-TMSAla-OH). Indeed, JMV2007 showed a drastic increase in affinity for both receptors ($IC_{50}$ of 0.02 nM for hNTS1 and 0.26 nM for hNTS2), leading to a NT(8-13) analog with one of the highest affinity described to date [61]. Finally, we combined the incorporation of $Sip^{10}$ and $TMSAla^{13}$ with the reduced amine bond which leads to JMV5296. While this compound showed a reduction of affinities at both NTS1 and NTS2, an increase in selectivity toward NTS2 (> 25-fold) was observed, with an extended resistance to protease degradation with a half-life of more than 20 hours. JMV5296 may thus represent an interesting compound for non-opioid pain management.

### *3.3 Central delivery of NT(8-13) analogs with a reduced amine bond reduces acute pain*

We first used the tail-flick acute pain paradigm to evaluate the antinociceptive response of NT(8-13) analogs bearing a reduced amine bond to thermal stimuli. Each analog was injected intrathecally (i.t.) at a dose of 60 μg/kg in male Sprague-Dawley rats and the tail withdrawal latency was evaluated over 40 min after spinal delivery (**Fig. 1A**). We found that all tested NT(8-13) analogs were able to trigger an antinociceptive response in this pain model. JMV449 displayed a transient antinociceptive response with a maximum at 10 min after injection before returning rapidly to basal level at 20 min. Maximal antinociceptive responses were observed after 20 min for JMV5206 and JMV5296 while it was delayed to 30 min for JMV5170. Despite the drastic loss of its binding affinities at both NT receptors, it is interesting to note that JMV5170 is still able to increase the response latency to thermal stimuli, in the same range as the other more potent NT(8-13) analogs.

We then calculated the area under the curve (AUC) for each analog to have a better understanding of the antinociceptive action over time (**Fig. 1B**). A similar increase of AUC was observed for JMV5170, JMV5206 and JMV5296, when compared to the saline-treated group. Despite showing a transient increase in tail withdrawal latency, JMV449 did not significatively increased AUC, which might reflect its short plasma half-life. Their analgesic responses to acute heat pain are comparable to non-selective NT(8-13) analogs that do not carry reduced bond (i.e. JMV2009 [36] and JMV2007 [61]), suggesting that in the case of multiple chemical modifications, the addition of the reduced bond did not confer a better efficacy in this pain paradigm. Interestingly, the response of these new NT derivatives was comparable to the one produced by equimolar dose of i.t. morphine, used here as the opioid reference analgesic (**Fig. 1**).

### 3.4 JMV5296 elicits moderated hypotensive response and no hypothermia

Before further consideration in more disease-relevant pain models, we evaluated the ability of these NT(8-13) analogs to produce hypotension and hypothermia, two known NTS1-related physiological effects [20]. We first evaluated the hypotensive action of these compounds after systemic administration. Arterial blood pressure was measured after i.v. injection of NT(8-13) analogs and reported as the difference in mean arterial blood pressure (ΔMABP) from the baseline recorded before injection (**Fig. 2A**). At a dose of 0.1 mg/kg, i.v. delivery of NT(8-13) induced a triphasic blood pressure response, a first short-lasting depressor phase of -30 mmHg followed by a pressor phase returning rapidly to baseline saline values and finally a third depressor phase leading to a robust and persistent drop in blood pressure (**Supplementary Fig. S1**) [62]. At a 10-fold lower dose (0.01 mg/kg), NT(8-13) did not produce this characteristic triphasic changes in blood pressure, the last sustained hypotensive phase having disappeared. At the dose of 0.01 mg/kg, JMV449 and JMV5206 were more potent than NT(8-13) in reducing blood pressure, producing the same triphasic profile of NT(8-13) at 0.1 mg/kg, with a sustained and maximal ΔMABP of -40 to -50 mmHg (**Fig. 2A**). At equimolar dose, JMV449 and JMV5206 were then as effective as the NTS1-selective agonist PD149163 (H-LysΨ[$CH_2NH$]Lys-Pro-Trp-Tle-Leu-OEt) in inducing hypotension (**Supplementary Fig. S1**) [53]. The high potency of these NTS1 analogs JMV449 and JMV5206 to induce drop in blood pressure is probably related to their high affinity for NTS1 as well as to their higher resistance to proteases. At the dose of 0.01 mg/kg, JMV5170 and JMV5296 caused a smaller first drop in blood pressure with a maximum reaching of -20 mmHg. As observed with NT(8-13) at the dose of 0.01 mg/kg, JMV5170 and JMV5296 did not induce the robust second hypotensive response, they rather provoked a sustained hypertensive phase, which is likely due to compensation mechanisms of the vascular system in response to the first drop (i.e., a release of catecholamines from the adrenal medulla), as described previously [62-66]. The remaining first hypotensive phase observed following administration of JMV5170 and JMV5296 can be attributed to their ability to bind NTS1, as the highly NTS2-selective ligand JMV431 fails to induce any drop of blood pressure at the same dose or at a 10-time higher dose (**Supplementary Fig. S1**). Hypotension mediated by NTS1, can be triggered after an i.v. injection of a non-selective neurotensinergic compound but also after i.t. administration at higher doses. As previously shown by [67], an i.t. administration of at least 75 μM of NT is required to induce a change in blood pressure. For the purpose of comparison, a dose of 60 μg/kg of NT(8-13) analog represents an i.t.

injection of less than 500 nM (for a rat weighing 300 g). Thus, when using spinal injections of NT(8-13) analogs, blood pressure lowering should not be a problem as long as the dose is not escalated.

Activation of the NTS1 receptor subtype has also been shown to produce mild hypothermia and to improve the neurological outcomes after ischemic stroke or traumatic brain injury [68-70]. We thus evaluated the effects of these NT(8-13) analogs on hypothermia after i.t. delivery at a dose of 30 µg/kg. Variation in body temperature was assessed every 10 min for up to 1 hour after i.t. administration and reported as the variation of temperature (Δ Body Temperature) from the baseline before injection (**Fig. 2B**). Our results showed that NT(8-13) and JMV449 had no effect on body temperature, compared to saline-treated animals while JMV5170 and JMV5206 induced a strong and sustained decrease of body temperature of more than 2°C for up to 1 hour. These results are consistent with previous findings showing that the stability of NT(8-13) analogs is a key factor driving hypothermic response and that the increased stability can even compensate for a reduction in binding affinity at NTS1 [39, 71]. Indeed, NT(8-13) and JMV449, which exhibit a plasma half-life of less than 10 min did not induce hypothermia. Accordingly, the compounds with extended plasma stability, JMV5170 and JMV5206, elicited important and sustained hypothermia in the same way as the NTS1-selective agonist PD149163 (H-LysΨ[CH$_2$NH]Lys-Pro-Trp-Tle-Leu-OEt), a NT(8-13) analog that also bears a reduced amine bond to improve its resistance to proteases [72] (**Supplementary Fig. S1**). Interestingly, despite its high metabolic stability, JMV5296 failed to induce any drop of core body temperature. This might be explained by its low affinity for NTS1 and enhanced selectivity toward NTS2. Accordingly, JMV431, a NTS2-selective compound (Boc-Arg-Arg-Pro-Tyr-(ψ(CH2-NH))-Ile-Leu), failed to induce hypothermia after i.t. injection (**Supplementary Fig. S1**) [73, 74]. Based on the concept of ligand bias signaling [75], we can also hypothesize that JMV5296 is able to discriminate between hypothermia and analgesia. This idea is supported by the demonstration that some NT(8-13) analogs, such as NT27 (Arg-D-Orn-Pro-Tyr-Ile-Leu) and NT77L (Arg-D-Orn-Pro-L-*neo*-Trp-Tle-Leu) acting at NTS1 are able to part between hypothermia and analgesia [76, 77].

Altogether, our characterization of this series of NT(8-13) analogs in acute thermal pain and for adverse effects revealed JMV5296 as a fairly NTS2-selective ligand with a prolonged plasma half-life and an improved benefit/risk ratio. These results prompted us to further characterize JMV5296 in more challenging pain paradigms.

### *3.5 JMV5296 alleviates tonic and chronic inflammatory pain*

Tonic pain induced by the injection of formalin into the rat hind paw is extensively used to investigate the ability of novel analgesics to reduce persistent inflammatory pain. Tonic pain models have been acknowledged to be of greater relevance to clinical pain than acute phasic pain models [47, 78]. This model has been previously used to characterize neurotensinergic compounds for their analgesic properties [28, 29]. Therefore, we evaluated the effect of an i.t. injection of JMV5296 at a dose of 60 µg/kg in the formalin-induced pain model and compared its analgesic effectiveness to equimolar dose of either NT(8-13) or morphine.

The injection of formalin in the hind paw of saline-treated animals produces a biphasic nociceptive response [79]. Although the first phase of nociceptive behaviors (acute phase, 0 to 9 min) reflects the direct activation of peripheral nociceptors, the second phase, observed between 21 to 60 min, results from peripheral inflammatory mediators, excitation of spinal dorsal horn neurons and central sensitization [47, 49, 78]. As compared to the saline-treated rats, both NT(8-13) and JMV5296 showed no significant reduction of the pain-related behaviors during the acute phase (**Fig. 3A**). However, in contrast to NT(8-13), JMV5296 exerted a sustained analgesic effect during the inflammatory phase, reducing the nociceptive behaviors by almost 85% when compared to the saline group (**Fig. 3B**). No motor adverse effects or altered consciousness were observed in rats treated with JMV5296. Importantly, JMV5296 was as effective as equimolar dose of i.t. morphine to reduce the formalin-induced nociceptive behaviors.

The analgesic response of JMV5296 in the formalin pain test is comparable to the response elicited by an i.t. injection of either JMV2007 or JMV2009, two non-selective NT agonists [36, 61]. Nevertheless, in contrast to JMV2007, JMV5296 did not decrease the nociceptive behaviors in the acute phase. When compared to the highly selective NTS2 ligand JMV431, the analgesic response elicited by JMV5296 in this persistent pain paradigm is greater than the one of JMV431 at the same dose of 60 µg/kg [28], indicating that a fairly selective NTS2 agonist can be of high interest. Indeed, it has been shown that NTS2 might not be involved during the first half of the second inflammatory phase [80] thus, the weak affinity of JMV5296 for NTS1 seems to be required to achieve a complete analgesic relief during the full inflammatory phase.

Finally, the analgesic potential of JMV5296 was evaluated in a chronic inflammatory pain model induced by intraplantar injection of the Complete Freund Adjuvant (CFA), creating an inflammatory chronic condition and the development of mechanical allodynia [81]. We evaluated the anti-allodynic efficacy of an i.t. injection of either NT(8-13) or JMV5296 at a dose of 60 µg/kg on days 3 and 14 after CFA injection, allowing us to assess their analgesic potential on the early inflammatory phase (3 days post-CFA) and after the development of chronic inflammatory pain (14 days post-CFA). The baseline paw withdrawal threshold (PWT) was determined before CFA injection. PWT was also recorded on the ipsilateral paw at days 3 and 14 post-CFA before i.t. injection of JMV5296 or saline. PWT was then determined at 15, 30, 60, and 90 min after spinal drug administration. Development of mechanical allodynia was comparable at days 3 and 14 post-CFA with a PWT around 4 g (**Fig. 4A, C**). At days 3 and 14 post-CFA, JMV5296 induced a significant and sustained reduction of mechanical allodynia characterized by the increase of PWT at 15- and 30-min post-injection followed by a slow return to the basal PWT at 60 min after spinal delivery (**Fig. 4A, C**). This reduction of mechanical allodynia can be better illustrated using the % of anti-allodynic effect at the maximal analgesic time point (i.e. 15 and 30 min for day 3 or 14, respectively). At day 3, JMV5296 produced a 52% reversal of allodynia, as compared to 59% at day 14 (**Fig. 4B, D**). In the same experimental pain paradigm, NT(8-13) was only effective in the early inflammatory phase.

Chronic inflammatory pain, which usually requires opioid dose escalation, is a debilitating condition often resulting in patient non-adherence to treatment [82]. Furthermore, this type of pain has multiple negative impacts, affecting patient's activities, deteriorating quality-of-life and severely increasing the apparition of comorbid conditions, such as sleep, anxiety, and depressive disorders [83, 84]. Compared to other chronic pain conditions like neuropathic pain, JMV5296 is as potent as other NT analogs like JMV431 and JMV2009 and even more potent than morphine or tricyclic antidepressants [36, 85, 86]. Thus, the strong anti-allodynic effects of JMV5296 could lead to an improvement in the management of health and quality-of-life outcomes. We could therefore suggest that reliable and effective pain relief in persistent and chronic inflammatory conditions could be reached using NT(8-13) analogs, acting as JMV5296, or by combining neurotensinergic drugs with opioid therapy, thus allowing the use of smaller doses and limiting undesired effects. Accordingly, combination therapy using NT analogs and morphine, or fentanyl, has already been found to be effective for pain relief and mitigating NT- and opioid-related side effects [87, 88]. Finally, these results demonstrated that JMV5296 induced a strong analgesic effect in both persistent and chronic inflammatory pain paradigms and that preferring NTS2-selective NT(8-13) analogs may represent an interesting alternative to limit the use of opioids and increase pain relief in chronic pain patients.

## 4 Conclusion and perspectives

The present study reports the characterization of a series of NT(8-13) analogs harboring a reduced amine bond and additional substitutions with unnatural silylated amino acids to give rise to metabolically stable and powerful analgesic pseudopeptide compounds. Incorporation of a reduced amine bond between Lys$^8$-Lys$^9$, Sip in position 10 and a TMSAla in position 13 of NT(8-13) resulted in the generation of JMV5296. These modifications produced a fairly NTS2-selective analog with extended resistance to proteolytic degradation which displays antinociceptive properties in acute, tonic and chronic pain models without inducing hypothermia. This study provides further evidence for NTS2 as a relevant target for pain modulation, thus offering a non-opioid option for the treatment of chronic pain.

However, one of the biggest challenges remaining in the development of NT(8-13) peptide analogs as potential analgesic drugs is the achievement of proper CNS penetration. Indeed, JMV5296 does not possess the characteristics that are favorable for crossing the blood-brain barrier (BBB) following systemic administration. We successfully applied a Trojan horse strategy to bypass the BBB and increase penetration of NT peptides using conjugation of NT with a brain-penetrant peptide ligand of LRP1 receptor (An2) [89]. Thus, conjugating JMV5296 with An2 is a strategy under investigation to generate brain-penetrant NTS2-selective analogs.

**CRediT authorship contribution statement**

**Mélanie Vivancos:** Methodology, Investigation, Formal analysis, Writing - original draft, Visualization. **Roberto Fanelli:** Resources. **Élie Besserer-Offroy:** Validation, Writing - review & editing. **Sabrina Beaulieu:** Methodology, Investigation. **Magali Chartier:** Methodology, Investigation. **Martin Resua-Rojas:** Investigation. **Christine E. Mona:** Resources. **Santo Previti:** Resources. **Emmanuelle Rémond:** Resources. **Jean-Michel Longpré:** Validation, Writing - review & editing. **Florine Cavelier:** Validation, review & editing, Supervision, Funding acquisition. **Philippe Sarret:** Validation, Writing - review & editing, Supervision, Funding acquisition.

**Florine Cavelier** and **Philippe Sarret** contributed equally to the supervision of this work and are designated to handle any correspondence and material requests. All authors have given approval to the final version of the manuscript.

**Conflict of interest**

The authors declare that they have no known competing financial interests or personal relationships which have, or could be perceived to have, influenced the work reported in this article.

**Acknowledgments**

The authors thank Professors Éric Marsault and Pedro D'Orléan-Juste (Dept. Pharmacology-Physiology, Université de Sherbrooke) for allowing them to use, the UPLC/MS instrument for plasma stability assays and the use of Micro-Med transducer for the blood pressure measurements, respectively.

MV was supported by a PhD scholarship from the Institut de Pharmacologie de Sherbrooke (IPS) and Centre d′Excellence en Neurosciences de l′Université de Sherbrooke (CNS). Funding from Montpellier University (FC for postdoctoral fellowship to RF) is also acknowledged. ÉBO is the recipient of a Fond de recherche du Québec – Santé (FRQ-S, 255989) and a Canadian Institute of Health Research (CIHR, MFE-164740) research fellowships. PS holds a Tier 1 Canada Research Chair in Neurophysiopharmacology of Chronic Pain and is a member of the FRQ-S-funded Québec Pain Research Network.

**Funding sources**

This research was supported by a grant from the Canadian Institutes for Health Research [grant number FDN-148413].

**Table**

**Table 1. Binding affinities and plasma stability of NT(8-13) analogs.** Bold indicates synthetic changes from native peptide NT(8-13). Data are expressed as mean ± SEM [39].

| Compound | Sequence | Binding (nM) $K_i$ ± SEM | | NTS2/NTS1 Selectivity | Plasma half-life |
| --- | --- | --- | --- | --- | --- |
| | | hNTS1 | hNTS2 | | |
| **NT(8-13)** | H-Arg-Arg-Pro-Tyr-Ile-Leu-OH | 1.65 ± 0.06 | 2.29 ± 0.16 | 0.72 | 0.98 ± 0.08 min |
| **JMV438** | H-Lys-Lys-Pro-Tyr-Ile-Leu-OH | 4.00 ± 0.35 | 1.12 ± 0.23 | 3.57 | 1.57 ± 0.27 min |
| **JMV449** | H-LysΨ[**CH$_2$NH**]Lys-Pro-Tyr-Ile-Leu-OH | 2.02 ± 0.80 | 0.29 ± 0.08 | 6.96 | 8.37 ± 2.02 min |
| **JMV5170** | H-LysΨ[**CH$_2$NH**]Lys-**Sip**-Tyr-Ile-Leu-OH | 296 ± 51 | 133 ± 33 | 2.23 | 22.1 ± 1.9 h |
| **JMV5206** | H-LysΨ[**CH$_2$NH**]Lys-Pro-Tyr-Ile-**TMSAla**-OH | 2.45 ± 0.17 | 0.54 ± 0.11 | 4.53 | 2.13 ± 0.19 h |
| **JMV5296** | H-LysΨ[**CH$_2$NH**]Lys-**Sip**-Tyr-Ile-**TMSAla**-OH | 610 ± 31 | 23.7 ± 4.6 | 25.4 | 20.6 ± 4.15 h |

**Figure Legends**

**Fig. 1. Antinociceptive action of NT(8-13) analogs with a reduced amine bond in acute thermal pain model. (A)** Time-dependent antinociceptive effect of NT analogs (60 µg/kg) or morphine at equimolar dose (25 µg/kg) on tail-flick latencies in rats. Baseline latencies were taken three times before acute i.t. injection. Latencies were determined every 10 min for up to 40 min following drug administration. *** $P < 0.001$ (vs. Saline) in a two-way ANOVA followed by Bonferroni's post hoc test. **(B)** Area under the curve (a.u., arbitrary units) determined for the entire duration of the tail-flick test (40 min) for each analog ** $P < 0.01$ and *** $P < 0.001$ (vs. Saline) in a one-way ANOVA followed by Dunnett's post hoc test for multiple comparisons. Each data point represents the mean ± S.E.M. n = 5-8 rats per group.

**Fig. 2. Effects of NT analogs on NTS1-mediated adverse effects. (A)** Effect of NT analogs on mean arterial blood pressure (ΔMABP) recorded on anesthetized rats after i.v. injection at a dose of 0.01 mg/kg. Each data point represents the mean ± S.E.M. n = 5-8 rats per group. **(B)** Hypothermia (ΔBody Temperature) induced by acute i.t. injection of saline or NT(8-13) analogs (30 µg/kg). Baseline body temperature assessment was performed before i.t. injection. A two-way ANOVA followed by Bonferroni's post hoc test was performed. * $P < 0.05$; ** $P < 0.01$; *** $P < 0.001$, as compared to saline-treated rats. Datapoints represent mean ± S.E.M. of determinations made in 5-8 rats.

**Fig. 3. Effects of central administration of JMV5296 on tonic persistent pain. (A)** Reduction of the nocifensive behaviors after i.t. injection of NT(8-13) (60 µg/kg), JMV5296 (60 µg/kg) or morphine at equimolar dose (25 µg/kg) in the formalin-induced tonic pain model. * $P < 0.05$; ** $P$

< 0.01; *** *P* < 0.001 (vs. Saline) in a two-way ANOVA followed by a Bonferroni's post hoc test. **(B)**. Cumulative nociceptive response expressed as AUC was measured during the total duration of the formalin test (0 – 60 min). * *P* < 0.05 and ** *P* < 0.01 (vs. Saline) in a one-way ANOVA followed by Dunnett's post hoc test for multiple comparisons. Each symbol represents the mean ± S.E.M. of determinations made in 5 rats.

**Fig. 4. Effects of JMV5296 on mechanical allodynia provoked by CFA injection.**
**(A, C)** Paw withdrawal thresholds were assessed with a manual von Frey hair at day 3 (**A**) and day 14 (**C**) after CFA injection. Baseline (BL) withdrawal thresholds were determined for all rats prior to CFA injection. Acute i.t. injection of JMV5296 or NT(8-13) (60 µg/kg) at days 3 and 14 post-CFA, effectively reduces the mechanical hypersensitivity. * *P* < 0.05; ** *P* < 0.01; *** *P* < 0.001 (vs. Saline) in a two-way ANOVA followed by a Bonferroni's post hoc test. Arrows indicate time of injection of JMV5296 or NT(8-13) **(B, D)** Percentage of anti-allodynia of NT(8-13) and JMV5296 at day 3 (**B**) and day 14 (**D**). JMV5296 was effective to attenuate the development of mechanical allodynia. * *P* < 0.05; *** *P* < 0.001 (vs. Saline) in a one-way ANOVA followed by Dunnett's post hoc test for multiple comparisons. Each datapoint represents mean ± SEM of 6-8 animals per group.

**Figures**

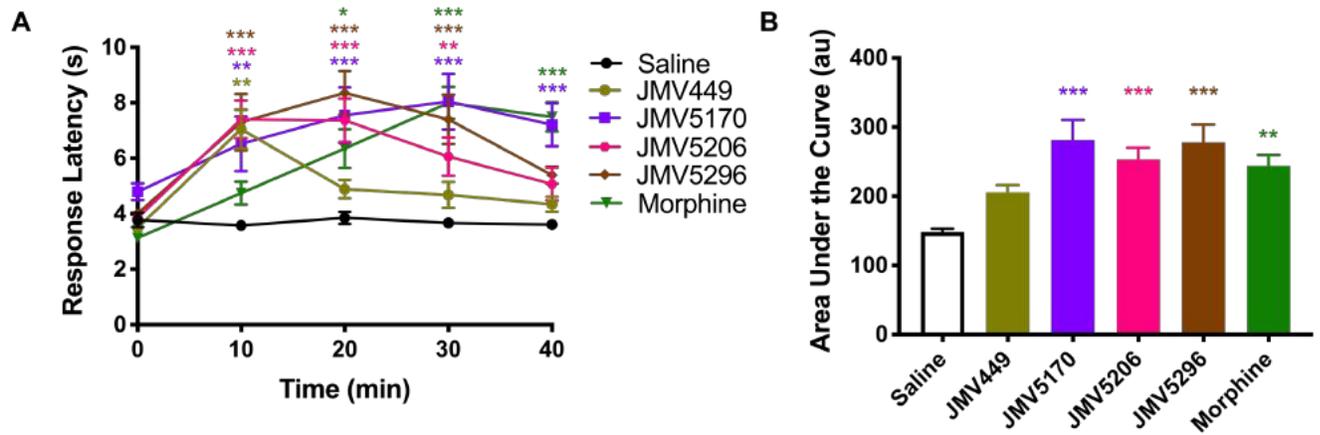

**Fig. 1.**

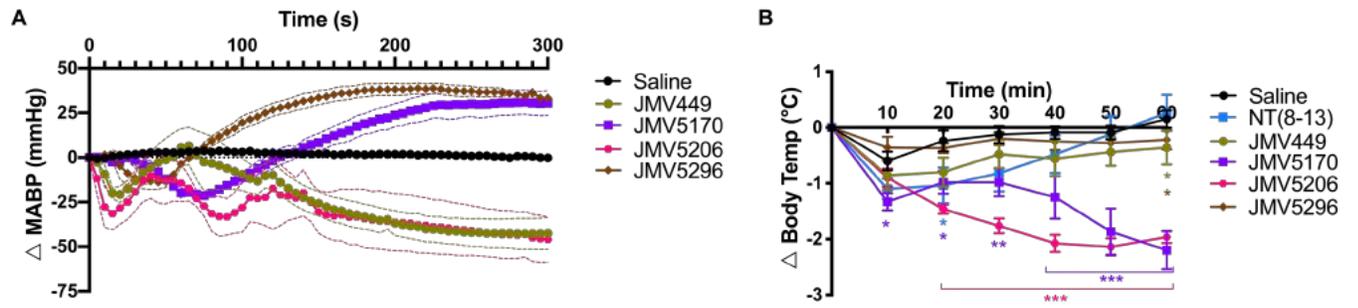

**Fig. 2.**

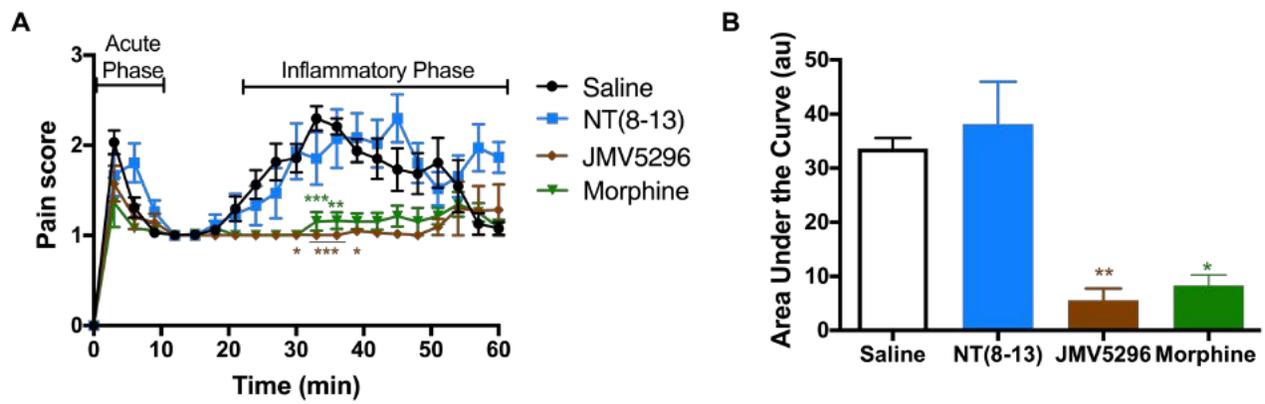

**Fig. 3.**

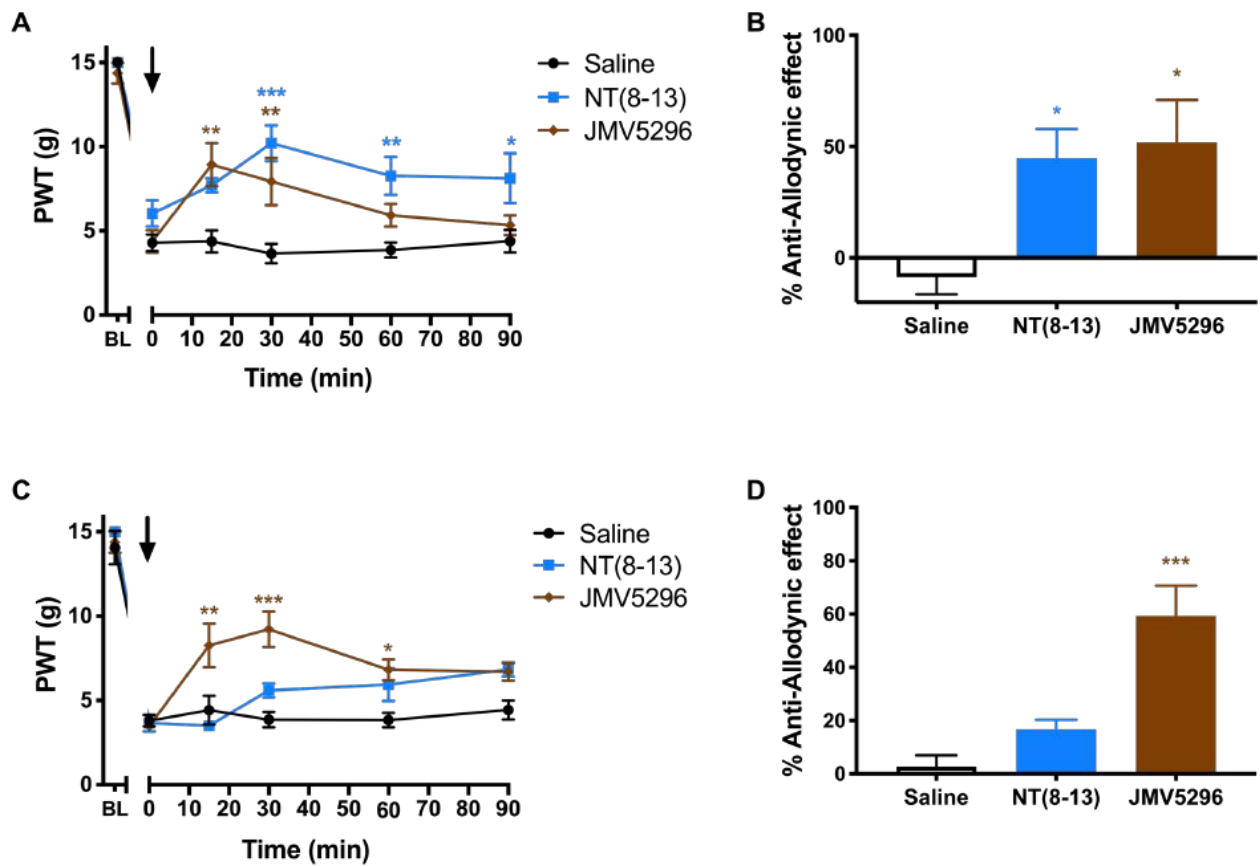

**Fig. 4.**

**Graphical abstract**

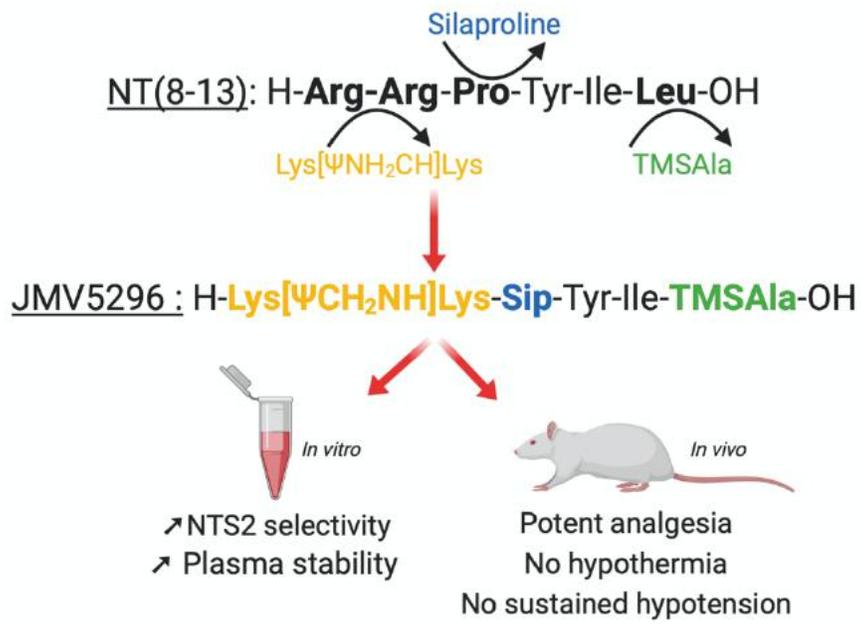